\documentclass[twocolumn,10pt]{IEEEtran}%
\usepackage[utf8]{inputenc}
\usepackage{amsfonts}
\usepackage{amsmath}
\usepackage{multicol}
\usepackage{amssymb}
\usepackage{graphicx}
\usepackage{mathtools}
\usepackage{xcolor}
\usepackage{ntheorem}
\usepackage{algorithm}
\usepackage{algorithmicx}
\usepackage{algpseudocode}
\usepackage{setspace}
\usepackage{listings}
\usepackage{cite}
\usepackage{url}
\usepackage{enumitem}

\algnewcommand\algorithmicoutput{\textbf{Output:}} 
\algnewcommand\Output{\item[\algorithmicoutput]}
\algnewcommand\algorithmicinput{\textbf{Input:}} 
\algnewcommand\Input{\item[\algorithmicinput]}

\begin{document}
\title{ Improved Coherence Index-Based Bound in Compressive Sensing} 

\author{
Ljubi\v{s}a~Stankovi\'{c},~\IEEEmembership{Fellow,~IEEE,} Milo\v{s}~Brajovi\'{c},~\IEEEmembership{Member,~IEEE,}
Danilo Mandic,~\IEEEmembership{Fellow,~IEEE,} Isidora~Stankovi\'{c},~\IEEEmembership{Member,~IEEE,}
Milo\v{s} Dakovi\'{c},~\IEEEmembership{Member,~IEEE},
\thanks{
L. Stankovi\'{c}. M. Brajovi\'{c}, I. Stankovi\'{c}, and M. Dakovi\'{c} are with the  University of
Montenegro, Podgorica, Montenegro. D. Mandic is with the Imperial College London, London, United Kingdom.
Contact e-mail: ljubisa@ac.me}
}
\maketitle

\begin{abstract}
Within the compressive sensing paradigm, sparse signals can be reconstructed based on a reduced set of measurements. The reliability of the solution is determined by its uniqueness. With its mathematically tractable and feasible calculation, the coherence index is one of very few CS metrics with considerable practical importance. 
In this paper, we propose an improvement of the coherence-based uniqueness relation for the matching pursuit algorithms. Starting from a simple and intuitive derivation of the standard uniqueness condition, based on the coherence index, we derive a less conservative coherence index-based lower bound for signal sparsity. The results are generalized to the uniqueness condition of the $l_0$-norm minimization for a signal represented in two orthonormal bases.
\end{abstract}

\section{Introduction and Basic CS Setting}

Compressive Sensing (CS) is a field that provides a rigorous framework for efficient data acquisition \cite{Intro,donoho2006,Tutorial,MP2,OMP,GSH1}. Examples include applications that rest upon reliable sensing from the lowest possible number of measurements, such as the recovery of sparse signals from vastly reduced sets of measurements and practical solutions in critical cases when some measurements are physically unavailable or heavily corrupted by disturbance.

Within the CS theory, several approaches have been established to reconstruct a sparse, $N$-dimensional vector, $\mathbf{X}$, from a reduced $M$-dimensional set of measurements, $\mathbf{y}$. The main concern in the reconstruction is to provide the conditions for a unique solution. Several frameworks for establishing the conditions for a unique solution are developed. The most important ones rely on the restricted isometry property (RIP) and the coherence index. While the RIP-based approach provides theoretically well-founded conditions, the main problem is in its computational feasibility \cite{Intro,donoho2006,DoL1L0,Tutorial0}. Namely, the RIP constant calculation is even more computationally demanding than the direct combinatorial solution of the CS problem itself. The coherence index-based condition is simple and computationally efficient. Its main disadvantage is that the reconstruction conditions based on this metric are quite pessimistic \cite{MP2,OMPCI}. 

Here, we will present an approach to alleviate this deficiency of the coherence index approach, introducing a computationally simple improved bound for the uniqueness relation based on the coherence index. The approach will be applied to a signal representation in two bases \cite{elad-bases}, being used for the derivation of the general sparsity bounds when the $\ell_0$ and $\ell_1$ minimizations are used to solve a CS problem.

\subsection{Definitions and Notation}

A sequence  $\{X(k)\}$, $k=0,1,\dots,N-1$ is referred to as a sparse sequence if the number, $K$, of its nonzero elements, $X(k)\ne 0$, is much smaller than its total length, $N$, that is,
$$X(k) \ne 0 \textrm{ for } k \in \{k_1,k_2,\dots,k_K\}, \,\, K \ll N.$$

A linear combination of elements of $X(k)$, given by
\begin{gather}
	y(m)=\sum_{k=0}^{N-1} a_m(k)X(k), \label{MeasDef}
\end{gather}  
is called a measurement, with the weights  denoted by $a_m(k)$.  

The above set of the measurements, $y(m)$, $m=0,1,\dots,M-1$, admits a vector/matrix form given by 
\begin{gather}
	\mathbf{y}=\mathbf{A}\mathbf{X},  \label{mmmess}
\end{gather}  
where $\mathbf{y}=[y(0),\,y(1),\dots,y(M-1)]^T$ is an $M\times 1$ column vector, $\mathbf{A}$ is an $M\times N$ measurement matrix which comprises the weights $a_m(k)$ as its elements, and $\mathbf{X}$ is an $N\times 1$ sparse column vector with elements $X(k)$. 

Without loss of generality, we shall assume that the measurement matrix, $ \mathbf{A}$, is normalized so that the energy of its columns sums up to unity. Consequently, the diagonal elements of its symmetric Gram form, $\mathbf{A}^H \mathbf{A}$, are equal to $1$, where $\mathbf{A}^H$ is the complex conjugate transpose of $ \mathbf{A}$. 

The compressive sensing theory task is to reconstruct the $N$-dimensional $K$-sparse vector $\mathbf{X}$ from a set of $M$ measurements, $\mathbf{y}=\mathbf{A}\mathbf{X}$, with  $K \ll M <N$. There are several approaches to solve this problem (for reviews of these approaches see \cite{Tutorial0,Tutorial}). Here we will consider the orthogonal matching pursuit (OMP) approach \cite{Tutorial0,Tutorial,OMP,OMPCI}.  

\subsection{OMP Solution to the CS Paradigm}

A matching pursuit reconstruction algorithm is typically based on a two-step strategy: 
\begin{itemize}
	\item[] Step 1: Detect the positions of nonzero elements,
	\item[] Step 2: Recover the signal by exploiting the relations between  measurement matrix, $\mathbf{A}$, detected positions and the vector of measurements, $\mathbf{y}$.
\end{itemize}

It will be further shown that  the physically relevant conditions for the reconstruction are in fact related to challenges emerging in the first step of the presented methodology. Otherwise, if arbitrary positions of $K$ nonzero elements of $\mathbf{X}$ are known, meaning that $X(k) \ne 0$  for $k \in \{k_1,k_2,\dots,k_K\}$, then a reduced set of measurement equations will follow as
$$\mathbf{y}=\mathbf{A}_{MK}\mathbf{X}_K,$$
with $\textbf{A}_{MK}$ being an $M\times K$ dimensional sub-matrix of matrix $\mathbf{A}$, formed by keeping the columns corresponding to the positions $\{k_1,k_2,\dots,k_K\}$.  Unknown values $X(k)$, located at $k \in \{k_1,k_2,\dots,k_K\}$,  are here conveniently grouped into a $K \times 1$ vector $\mathbf{X}_K$. This system of $M$ equations and $K<M$ unknowns
has a solution in a Least Square (LS) sense, in form
\begin{equation}\mathbf{X}_K=(\mathbf{A}^H_{MK}\mathbf{A}_{MK})^{-1}\mathbf{A}_{MK}^H\mathbf{y}=\textrm{pinv}(\mathbf{A}_{MK})\mathbf{y}. \label{pinvSol}
\end{equation} 
A sufficient condition for this reconstruction with known positions is that the matrix $\mathbf{A}^H_{MK}\mathbf{A}_{MK}$  is regular. 
The more demanding condition is that the positions of the nonzero elements in the sparse vector are exactly determined. 

\textit{The reconstruction solution is exact if the positions  $\{k_1,k_2,\dots,k_K\}$ of nonzero elements in a $K$-sparse vector $\mathbf{X}$ are exactly  determined for any set  $\{k_1,k_2,\dots,k_K\}$ and if there exist at least $K$ independent measurements \cite{OMPCI,OMP}.} 

This means that the detection step of the OMP approach is crucial for the exact solution. The detection is based on the initial estimate, defined as the back-projection of the measurements, $\mathbf{y}$, to the measurement matrix, $\mathbf{A}$, in the form
\begin{equation}
	\mathbf{X}_0=\mathbf{A}^H \mathbf{y}=(\mathbf{A}^H \mathbf{A})  \mathbf{X}.
	\label{Initi}
\end{equation}
If $\mathbf{A}^H\mathbf{A}$ ensures that the largest $K$ elements of the initial estimate, $\mathbf{X}_0$ are positioned at exact $k \in \{k_1,k_2,\dots,k_K\}$, then the detection is performed by taking the positions of the highest magnitude elements in the initial estimate, which is followed by the reconstruction based on (\ref{pinvSol}).  The condition that $K$ elements in the initial estimate, $\mathbf{X}_0$ located at the positions of non-zero elements in the original sparse vector, $\mathbf{X}$ are larger than any other component in the initial estimate can be relaxed through an iterative procedure. 

In such methodology, being the basis of matching pursuit algorithms such as the OMP \cite{OMP}, in order to find the position, $k_1$ of the largest non-zero element in $\mathbf{X}_0$, it is required that its value is larger than any value at the original zero-valued element position. Upon detecting the position, $k_1$, and estimating the component value based on (\ref{pinvSol}), with $\mathbf{A}_K$ being formed based on $k_1$, the contribution of this component is removed from measurements vector, $\mathbf{y}$. The procedure is iteratively repeated for the remaining $(K-1)$ elements. If the uniqueness condition is satisfied for the exact detection of the position of the largest nonzero element, then this condition is satisfied for the remaining $(K-1)$-sparse problem (less restrictive problem, with the sparsity reduced from $K$ to $K-1$).
  
\section{Uniqueness of the OMP Reconstruction}

The uniqueness condition based on the coherence index can be formulated as follows.

\noindent \textit{The reconstruction of a $K$-sparse signal, $\mathbf{X}$, is unique if the coherence index,
	\begin{equation}
		\mu = \max_{\substack{k,l \\ k\neq l}} \left| \sum_{m=0}^{M-1}a_m(k)a^*_m(l) \right|, \label{Cohkoc}
	\end{equation}
	of the normalized measurement matrix, $\mathbf{A}$, satisfies \cite{donoho2006}
	\begin{equation}
		K < \frac{1}{2}\left(1+\frac{1}{\mu}\right), \label{Kkoc}
	\end{equation}
	The coherence index, $\mu$, is equal to the maximum absolute off-diagonal element of $\mathbf{A}^H \mathbf{A}$, while its diagonal elements are equal to $1$. 
	The condition in (\ref{Kkoc}) guarantees that the solutions obtained by minimizing the $\ell_0$-norm and $\ell_1$-norm produce the same common unique solution \cite{DoL1L0}. This condition guaranties unique solution produced by the OMP algorithm \cite{GSH,OMPCI,OMP,MP2,coher-mag}.
}

Although this criterion is commonly derived based on the support uncertainty principle \cite{DoL1L0,MP2} or Gershogorin disk theorem \cite{GSH},  the coherence index condition (\ref{Cohkoc}) follows also as a result of the analysis in the process of detection of positions of non-zero values in original vector $\mathbf{X}$ \cite{coher-mag}.

By definition, any measurement represents a linear combination of nonzero elements of the sparse vector $\mathbf{X}$, that is
$$y(m)=\sum_{i=1}^{K}X(k_i)a_m(k_i).$$
Furthermore, without loss of generality, it can be assumed that the largest element is $X(k_1)=1$, whereas the remaining nonzero elements do not take values greater than this value, $\left|X(k_i)\right|\le 1$, $i=2,3,\dots,K$. In that case, the initial estimate 
$$
X_0(k)=\sum_{i=1}^{K}X(k_i)\sum_{m=0}^{M-1}a_m(k)a^*_{m}(k_i)=\sum_{i=1}^{K}X(k_i) \mu(k,k_i)
$$
can be expressed, for the element at $k=k_1$, as follows
\begin{equation}X_0(k_1)=X(k_1)+\sum_{i=2}^{K}X(k_i) \mu(k,k_i),  \label{ini}
\end{equation}
where
$\mu(k,k_i)=\sum_{m=0}^{M-1}a_m(k)a^*_{m}(k_i)$. The maximum possible absolute value of $\mu(k,k_i)$ is then equal to the coherence index, that is, $\mu=\max_{k,k_i}|\mu(k,k_i)|$.

In the worst case scenario for the detection of the element at position $k_1$, the value of this element, $|X_0(k_1)|$ in (\ref{ini}) is maximally reduced by the term $\sum_{i=2}^{K}X(k_i) \mu(k,k_i)$. The maximally reduced coefficient  $|X_0(k_1)|$ takes the value
\begin{equation}
	\min|X_0(k_1)|=1-\sum_{i=2}^{K}|X(k_i) \mu(k,k_i)|=1-(K-1)\mu, \label{kminus1}
\end{equation}
assuming that all $K-1$ remaining elements $X(k_i)$
have the most unfavorable value, $X(k_i)=1$, whereas $|\mu(k,k_i)|=\mu$, for each $k_i\in\{k_1,k_2,\dots,k_K\}$. 

The maximum value of disturbance at the positions where the elements were originally zero-valued, $k \notin \{k_1,k_2,\dots,k_K\}=\mathbb{K}$ is equal to 
\begin{align}
	\max_{k, k\notin \mathbb{K}}|\sum_{i=1}^{K}X(k_i)\mu(k,k_i)|=K\mu. \label{kvalues}
\end{align}

In the worst case scenario, the exact and unique detection of the position of the largest element $X_0(k_1)$ is possible when its maximally degraded value, exceeds the maximal value of the disturbance
\begin{equation*}
	\min|X_0(k_1)|>\max_{k\ne k_i}|\sum_{i=1}^{K}X(k_i)\mu(k,k_i)|,
\end{equation*}
or equivalently,  $1-(K-1)\mu>K\mu$, producing (\ref{Kkoc}).

Upon successfully detecting, reconstructing, and removing the first  non-zero component in a sparse $\mathbf{X}$,
the same procedure and relations can be iteratively applied to the remaining ``deflated" signal which now exhibits a reduced 
$(K-1)$-sparsity level, thus guaranteeing an exact and unique solution.

\section{Improved Bound Derivation}

In the previous derivation of the  reconstruction relation  (\ref{Kkoc}), 
it has been assumed that $K$ maximum absolute values of $\mu(k,k_i)=\mu$ in (\ref{kvalues}) add up to form the disturbance. Moreover, it has been assumed that the component $X_0(k_1)$, that we aim to detect at a position $k_1$, is reduced by $K-1$ maximal values  of $\mu(k,k_i)=\mu$. This is, however, an overly pessimistic assumption, since even in the worst case scenario all the largest $2K-1$ values of $\mu(k,k_i)$, in general, may not be equal to $\mu$.

Actually, when the first maximum is taken $|\mu(k,k_i)|=\mu$, in the next sample  only  the second largest value of $|\mu(k,k_i)|$ can be taken.  Subsequently, only the third largest value of $|\mu(k,k_i)|$ can be taken, and so on.
To take this fact into account and derive a less conservative reconstruction bound, denote the sorted values of $ |\mu(k,k_i)|$ as
\begin{gather}s(p)=\mathrm{sort}_{k,k_i}\{|\mu(k,k_i)|\}, \label{SortS}\\
 k, k_i=1,2,\dots,N, \ \ p=1,2,\dots, N^2, \nonumber
\end{gather}
assuming a nonincreasing order, $ s(1)  \ge s(2) \ge\dots \ge s(N^2) $.
In the worst case scenario, instead of $2K-1$ values of $\mu$, now we can use the first $(2K-1)$ (largest) values of $s(p)$  to get
$$1>\sum_{p=1}^{2K-1}s(p)=(2K-1)\alpha_{\mathbf{A}}(2K-1),$$ 
instead of $1>(K-1)\mu+K\mu$, where $\alpha_{\mathbf{A}}(2K-1)$ is the mean value of the $(2K-1)$ largest values of $ |\mu(k,k_i)|$,   $$\alpha_{\mathbf{A}}(2K-1)=\frac{1}{2K-1}\sum_{p=1}^{2K-1}s(p)=\mathop{\mathrm{mean}}_{\substack {1\le p \le 2K-1}}{s(p)}.$$
The bound for the reconstruction now becomes
\begin{equation}K<\frac{1}{2}\left(1+\frac{1}{\alpha_{\mathbf{A}}(2K-1)}\right).\label{muavbound}
\end{equation}
This implicit inequality is easily solved by direct check, starting from $K=1$, and then increasing the value of $K$ by one, until the inequality holds \cite{coher-mag}. The procedure is stopped for the smallest $K$ when (\ref{muavbound}) does not hold.  

In the special case of an equiangular tight frame (ETF) measurement matrix, when the factor $|\mu(k,k_i)|=\mu$ is constant, then  $\alpha_{\mathbf{A}}(2K-1)=\mu$ and (\ref{Kkoc}) holds.
In all cases, for any measurement matrix $\mathbf{A}$, condition
 $\alpha_{\mathbf{A}}(2K-1)\le {\mu}$
holds. This means that a more optimistic bound for $K$ is obtained by (\ref{muavbound}) than the conventional CS bound in (\ref{Kkoc}). 
 
Furthermore, it will be shown that even a less conservative bound than that in (\ref{muavbound}) can be derived following some simple observations of the initial estimate calculation based on the Gram matrix $\mathbf{A}^H\mathbf{A}$.
 Recall that the value of the initial estimate, $X_0(k)$, at a non-zero position $k_1$ can be calculated using (\ref{ini}) at $k=k_1$.  
 
In the described worst possible scenario, the observed “largest” term $X(k_1)=1$ is maximally reduced. 
 This happens when $|\mu(k_1,k_i)|$ takes the largest possible values \textit{only within a row  with index $k_1$} of matrix $\mathbf{A}^H\mathbf{A}$.  If we sort into a nonincreasing order values of rows (for any $k_1$) and form
\begin{equation}
	s_1(p)=\mathrm{sort}{|\mu(k_1,l)|} \label{ss}
\end{equation} 
$l=1,2,\dots,N$, such that $s_1(1)\ge s_1(2)\ge\dots\ge s_1(N)$, then in the worst case scenario, $X(k_1)$ will be reduced for 
\begin{equation}
(K-1)\frac{1}{K-1}\sum_{p=1}^{K-1}s_1(p)=(K-1)\beta_{\mathbf{A}}(K-1) \label{b1}
\end{equation}
 that is, by the first $(K-1)$  coefficients $s_p(l)$, being in fact the largest possible $(K-1)$ values of $|\mu(k_1,l)|$ for any $k_1$, that is $\beta_{\mathbf{A}}(K-1)=\max_{k_1}\{ \mathrm{mean}_{1\le p \le K-1}{s_1(p)}\}$. Note that (\ref{b1}) is obtained based on the second part of (\ref{ini}). The fact that $X(k)=1$ for all $k\in \{k_1,k_2,\dots,k_K\}$ implies that each component equally contributes to this reduction. 


The largest possible disturbance value is obtained when absolute values of the elements $|\mu(k,k_i)|$ of matrix $\mathbf{A}^H\mathbf{A}$  take the largest values \textit{in the given row at the disturbance position} $k\notin \{k_1,k_2,\dots,k_K\}$ and are summed in phase. If we take into account the notation for sorted values in (\ref{ss}), then this accumulated disturbance value becomes
\begin{equation}
	K \gamma_{\mathbf{A}}(K)=K\frac{1}{K}\sum_{p=1}^{K}s_2(p)=K\mathop{\mathrm{mean}}_{\substack {1\le p \le K}}{s_2(p)} \label{b2}
\end{equation} 
having in mind that $|\mu(k,k_i)|$ takes $K$ largest values from \textit{one row} (excluding the values in the rows taken in $s_1(p)$), as opposed to the previous assumption that it takes the single largest value $\mu$ repeatedly $K$ times (which leads to the traditional coherence index bound).

The successful detection of component $X(k_1)$  will not be compromised if  the component, assuming its smallest possible value, is still larger than the largest value of the  disturbance at $k\notin \{k_1,k_2,\dots,k_K\}$ 
$$ 1-\sum_{i=2}^{K}|X(k_i) \mu(k,k_i)|< \sum_{i=1}^{K}X(k_i) \mu(k,k_i)$$
or, having in mind (\ref{b1}) and (\ref{b2})
\begin{gather*}
	1>(K-1)\beta_{\mathbf{A}}(K-1)+K\gamma_{\mathbf{A}}(K),
\end{gather*}
where
\begin{gather*}
	\beta_{\mathbf{A}}(K-1)=\mathop{\mathrm{mean}}_{\substack {1\le p \le K-1}}{s_1(p)} \text{ and } 
	\gamma_{\mathbf{A}}(K)=\mathop{\mathrm{mean}}_{\substack {1\le p \le K}}{s_2(p)}.
\end{gather*} 

\textit{The reconstruction of a K-sparse signal, $\mathbf{X}$, is exact and unique if the measurement matrix, $\mathbf{A}$, guarantees that the following condition is satisfied
\begin{gather}
	K<\frac{1+\beta_{\mathbf{A}}(K-1)}{\beta_{\mathbf{A}}(K-1)+\gamma_{\mathbf{A}}(K)}. \label{genKb}
\end{gather}
}

The three discussed sparsity bounds are related as
\begin{gather*}
	\frac{1+\beta_{\mathbf{A}}(K-1)}{\beta_{\mathbf{A}}(K-1)+\gamma_{\mathbf{A}}(K)} \ge \frac{1}{2}(1+\frac{1}{\alpha_{\mathbf{A}}(2K-1)}) \ge \frac{1}{2}(1+\frac{1}{\mu}).
\end{gather*} 
The equality holds for the ETF measurement matrices when 
$\beta_{\mathbf{A}}(K-1)=\gamma_{\mathbf{A}}(K)=\alpha_{\mathbf{A}}(2K-1)=\mu$.

 \section{Numerical Examples}
 
%
%

 The presented relations are tested on several numerical examples: a graph matrix, a measurement matrix of a Gaussian form,  partial DFT and DCT matrices, and an ETF form.

\begin{figure}
 	[ptb]
 	\begin{center}

 		 \includegraphics[trim=4cm 0 2cm 0,clip,width=3.5cm]{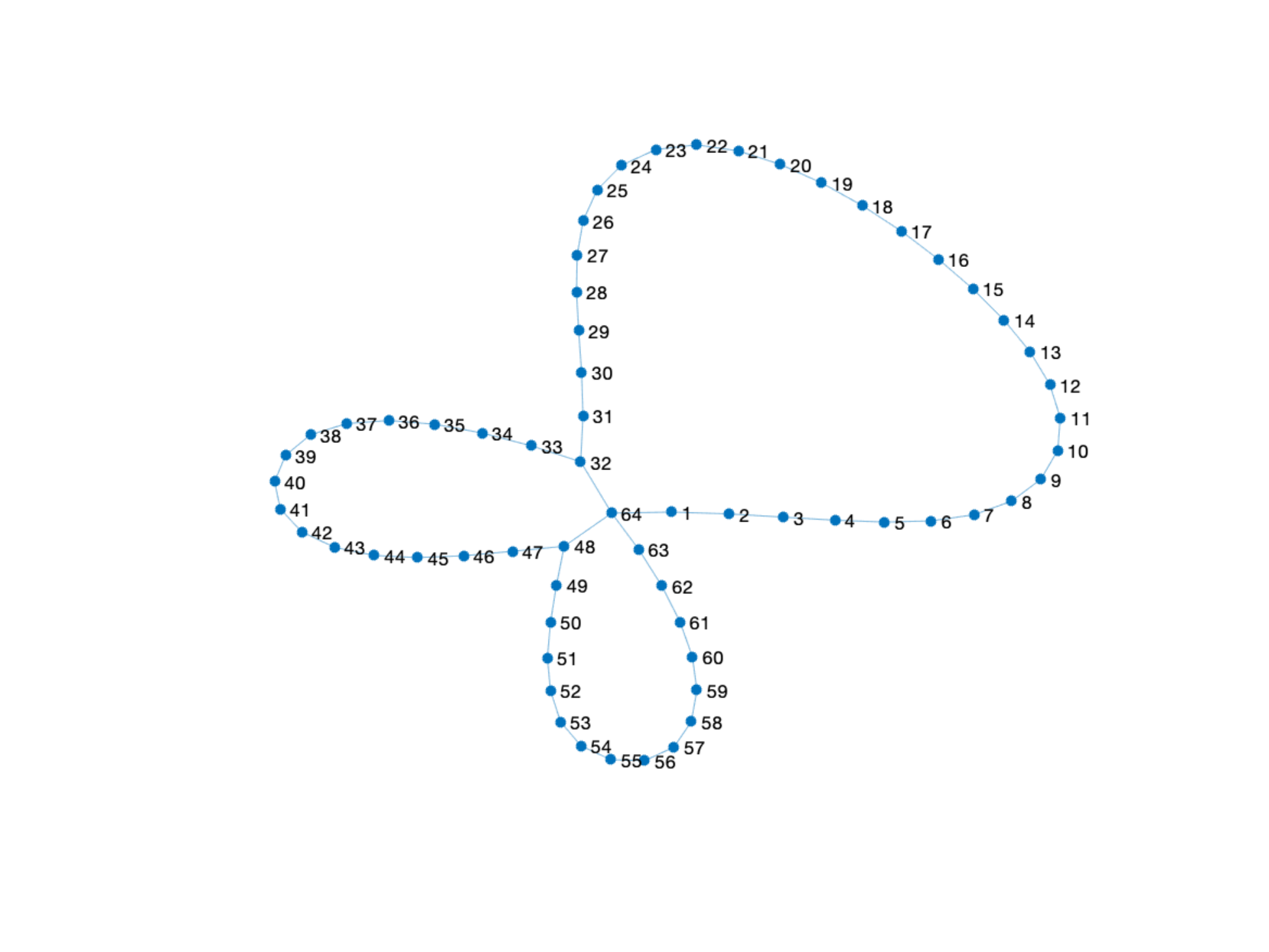}	
 		\ \ \ 	\includegraphics[scale=0.25]{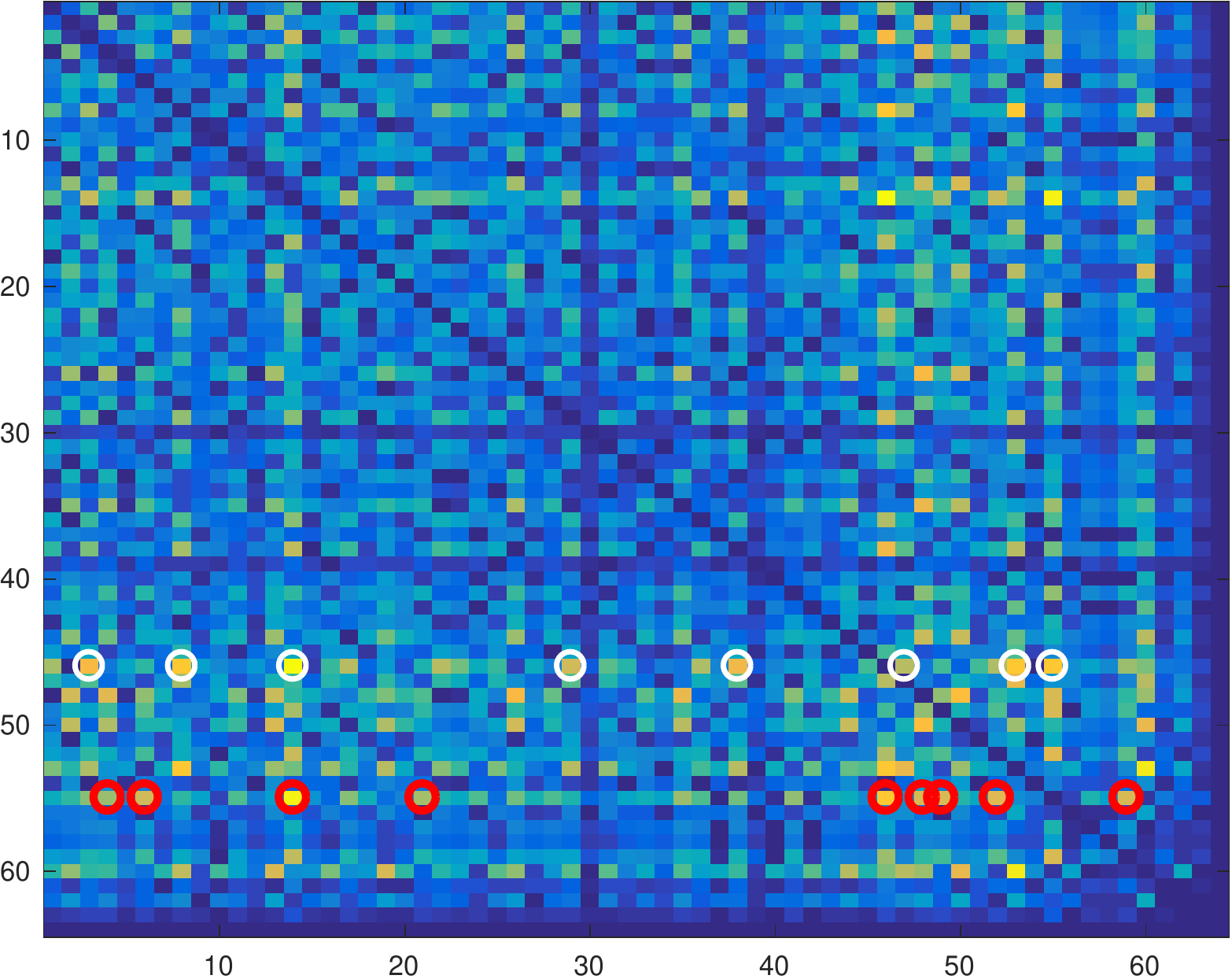}
 		
 		\caption{A graph (top) and the matrix $\mathbf{A}^H\mathbf{A}$. The elements of matrix $\mathbf{A}^H\mathbf{A}$ used for the calculation of $\beta_{\mathbf{A}}(K-1)$ are circled using a white line,  while its elements  used for the calculation of $\gamma_{\mathbf{A}}(K)$ are circled using a red line, with $K=9$, being the smallest $K$ when  inequality (\ref{genKb}) is not satisfied.}\label{graph_SQ}
 	\end{center}
 \end{figure}

 
 An unweighted and undirected graph is given in Fig. \ref{graph_SQ}(a). The graph Fourier transform (GFT) matrix, for the spectral representation, is defined by the eigenvectors of the graph Laplacian, as its columns \cite{GGR}. It has been assumed that the graph signal is $K$-sparse in the GFT domain and that the samples at vertices $n=21$ and  $n=38$ are missing.   The Gram matrix $\mathbf{A}^H\mathbf{A}$ of $\mathbf{A}$, corresponding to $M=62$ available samples of the GFT matrix, is shown in Fig. \ref{graph_SQ}(b). This matrix guarantees unique reconstruction for $K<6.8917$, $ K<7.4618$, and $K< 8.3118$, respectively, with the three presented approaches for the sparsity bound determination, given by relations (\ref{Kkoc}), (\ref{muavbound}), and (\ref{genKb}), respectively. We can conclude that the sparsity limit is improved from the maximal sparsity $K=6$, with (\ref{Cohkoc}), to $K=8$ using (\ref{genKb}). The results are statistically checked. It has been concluded that in $10^6$ random realizations for $K\le8$ all reconstructions were successful. 
 
 For a Gaussian measurement matrix we used $N=80$ and $M=70$. In 1000 realizations (with various Gaussian matrices) we obtained the mean value of the limit $K<1.6761$ with a standard deviation (SD) of $0.08$, while the presented method produced the mean value limit $K<2.3523$ with an SD of $0.16$. The same experiment with a $1000 \times 900$ measurement matrix produced $K<3.6175$ with an SD of $0.15$ and $K<4.3580$ with an SD of $0.07$, meaning that the sparsity of a certain exact reconstruction is improved from $3$ to $4$.
 
 For a partial DCT matrix, with dimension $128 \times 124$, we obtained the mean values  $K<9.7849$ and $K<12.1354$, with (\ref{Kkoc}) and (\ref{genKb}). The best case in $1000$ random realizations of the available samples, with improved bound, was $K<14.2238$.  
 
 For an $128 \times 124$ partial DFT matrix, we obtained the mean values $K<16.9068$ and $K<19.8323$, with (\ref{Kkoc}) and (\ref{genKb}). The best case in $1000$ random realizations of the available samples was $K<21.4307$ with (\ref{genKb}). For just 2 missing samples, the measurement matrix behaves close to an ETF and produced the limits for $K$ very close to $N/4=32$.  For $128 \times 112$ measurement matrix the best limit was $K<10.3770$. For half of the available samples, $M=N/2$, all limits drop toward the theoretically the worst case when no unique solution can be achieved. In $1000$ realizations the mean values are just above $3$, the best form of the measurement matrix produced the sparsity limit slightly above $4$, while for the  worst measurement matrix in these $1000$ realizations, the limit dropped to $K<1.9545$.   With a $128 \times 20$, corresponding to just $20$ available samples, the mean limits were  $K<1.6325$, and $K<2.2649$.   
 
 Finally, for an ETF matrix of dimension $18 \times 9$ all the presented limits were the same, as expected since the absolute values of the off-diagonal elements of $\mathbf{A}^H\mathbf{A}$ are the same. The common limit is $K<2.5616$.

\section{Generalization For Two Bases  and $\ell_0$-norm}

The presented framework can be used to generalize the results obtained analyzing the signal representation in two bases, as introduced in \cite{elad-bases}. This kind of signal representation was used to find the general sparsity bounds for the unique solutions, obtained using the $\ell_0$-norm and  the $\ell_1$-norm minimizations. 
Consider representations of  a unit energy signal, $x(n)$, in  two arbitrary  bases $u_k(n)$ and $v_l(n)$, with respective transformation elements $X(k)$ and $Y(l)$. Assume, as in \cite{elad-bases,donoho2006}, that the signal is sparse in these bases with sparsities $||\mathbf{X}||_0=K$ and $||\mathbf{Y}||_0=L$, and that the Parseval's theorem holds in both bases,  $||\mathbf{X}||^2_2=1$ and $||\mathbf{Y}||^2_2=1$. Form a function $L(n,k,l)=X(k)Y^*(l)u_k(n)v^*_l(n)$ as in \cite{graph-bound} such that 
\begin{gather*}
	\sum_{n}x(n)x^*(n)=\sum_{n}\sum_{k}\sum_{l}X(k)Y^*(l)u_k(n)v^*_l(n)=1,
\end{gather*}  
where  $k \in\{k_1,k_2,\dots,k_{K}\}$ and $l \in\{l_1,l_2,\dots,l_{L}\}$, then using Schwartz's inequality we get
\begin{gather*}
	1= \Big|\sum_{k}\sum_{l}X(k)Y^*(l)\sum_{n}u_k(n)v^*_l(n)\Big|^2 \le \\ \sum_{k}\sum_{l}|X(k)|^2|Y(l)|^2\sum_{k}\sum_{l}|\mu(k,l)|^2\le KL \frac{1}{KL} \sum_{p}s^2(p)
\end{gather*}  
where $s(p)$ is defined in (\ref{SortS}). Using the notation $\eta_\mathbf{A}(KL)=\frac{1}{KL}\sum_{k}\sum_{p}s^2(p)$  and $\sqrt{ab}\le(a+b)/2$, $a>0$, $b>0$, we get
\begin{gather*}
	\frac{1}{\eta_\mathbf{A}(KL)} \le KL =||\mathbf{X}||_0||\mathbf{Y}||_0 \le \Big(\frac{1}{2}(||\mathbf{X}||_0+||\mathbf{Y}||_0)\Big)^2 \\
\text{or \ \ \ \ \ }	||\mathbf{X}||_0+||\mathbf{Y}||_0\ge \frac{2}{\sqrt{\eta_\mathbf{A}(KL)}}.  
\end{gather*}
The solution of the $\ell _0$-norm minimization is unique if the sparsity, $	||\mathbf{X}||_0$, is smaller than half of the uncertainty bound
\begin{gather*}
	K < \frac{1}{\sqrt{\eta_\mathbf{A}(K^2)}}\ge \frac{1}{\mu}.
\end{gather*}
This relation can be used to derive improved coherence index-based conditions when the $\ell _0$-norm and $\ell _1$-norm minimization produce the same and unique solution \cite{elad-bases,donoho2006}.

\section{Conclusion}
A numerically efficient calculation of an improved coherence index-based sparsity bound is proposed. The calculation is demonstrated on a graph signals example and several commonly used measurement matrices. The results are generalized for the $l_0$-norm and two bases.

\end{document}